\documentclass[prl,twocolumn,superscriptaddress]{revtex4}
\usepackage{amsmath}
\usepackage{amsfonts}
\usepackage{amssymb}
\usepackage{graphicx}

\begin{document}

\title{Crucial Role of Quantum Entanglement in Bulk Properties of Solids}

\author{\v{C}aslav Brukner}
\affiliation{Institut f\"ur Experimentalphysik, Universit\"at Wien,
Boltzmanngasse 5, A--1090 Wien, Austria} \affiliation{The Blackett
Laboratory, Imperial College, Prince Consort Road, London, SW7 2BW,
United Kingdom}
\author{Vlatko Vedral}
\affiliation{The School of Physics and Astronomy, University of
Leeds, Leeds, LS2 9JT, United Kingdom} \affiliation{The Erwin
Schr\"{o}dinger Institute for Mathematical Physics, Boltzmanngasse
9, A--1090, Vienna, Austria}
\author{Anton Zeilinger}
\affiliation{Institut f\"ur Experimentalphysik, Universit\"at Wien,
Boltzmanngasse 5, A--1090 Wien, Austria} \affiliation{Institut f\"ur
Quantenoptik und Quanteninformation, \"Osterreichische Akademie der
Wissenschaften, Boltzmanngasse 3, A--1090 Wien, Austria}

\date{\today}

\begin{abstract}

We demonstrate that the magnetic susceptibility of strongly
alternating antiferromagnetic spin-1/2 chains is an entanglement
witness. Specifically, magnetic susceptibility of copper nitrate
(CN) measured in 1963~\cite{berger} cannot be described without
presence of entanglement. A detailed analysis of the spin
correlations in CN as obtained from neutron scattering
experiments~\cite{broholm} provides microscopic support for this
interpretation. We present a quantitative analysis resulting in the
critical temperature of 5K in both, completely independent,
experiments below which entanglement exists.
\end{abstract}


\maketitle

Entangled quantum systems can exhibit correlations that cannot be
explained on the basis of classical laws. Since the birth of quantum
theory, such correlations have been used to highlight a number of
counter-intuitive phenomena, such as the Einstein-Podolsky-Rosen
paradox~\cite{epr} or quantum non-locality~\cite{bell} - the
conflict between quantum mechanics and local realism as quantified
by violation of Bell's inequalities. In recent years entanglement
was realized to be a crucial resource that allows for powerful new
communication and computational tasks that are not possible
classically~\cite{nielsenchuang}.

The existence of quantum entanglement is generally not seen beyond
the atomic scale. Only very recently entanglement experiments were
realized with increasingly complex objects, either by entangling
more and more systems with each
other~\cite{ghz,4photon,5photon,polzik}, or by entangling systems
with a larger number of degrees of freedom~\cite{vaziri}. Moving
towards demonstration of entanglement at even larger scales will
tackle the question on limits on mass, size and complexity of
systems that still can contain entanglement. The usual arguments
against seeing entanglement on macroscopic scales is that large
objects contain large number of degrees of freedom that can interact
with environment thus inducing decoherence that ultimately lead to a
quantum-to-classical transition.

Remarkably, macroscopic entanglement in solids, that is,
entanglement in the thermodynamical limit of infinite large number
of constituents of solids, was theoretically predicted to exist even
at moderately high
temperatures~\cite{nielsen,arnesen,oconnor,osborne,vlatkosuperconducter}.
Recently, it was demonstrated that entanglement can even affect
macroscopic thermodynamical properties of
solids~\cite{oconnor,wang1,wang2,vlatko,caslav,ghosh}, such as its
magnetic susceptibility or heat capacity, albeit at very low
temperature (few milikelvin) and only for a special material system
- the insulating magnetic salt LiHo$_x$Y$_{1-x}$F$_4$~\cite{ghosh}.
This extraordinary result shows that entanglement can have
significant macroscopic effects.

Nevertheless, it is an open question whether or not there are
macroscopic thermodynamical quantities that are entanglement
witnesses for broader classes of solid state systems.
Entanglement witnesses are observables which have positive
expectation values for separable states and negative ones for
some entangled states~\cite{horodecki}. Finally, a direct
experimental demonstration of entanglement, in the form of a
determination of correlations between microscopic constituents of
solids, remains an experimental challenge.

Here we show that bulk properties of solids can be entanglement
witnesses. We use already published experimental results of both
microscopic structure and macroscopic properties of the spin-1/2
alternating bond antiferromagnet CN (Cu(NO$_3$)$_2$2.5D$_2$O), to
demonstrate the existence of entanglement in this type of solid. In
the first approach we analyse experimental results of neutron
scattering measurement of CN obtained in 2000~\cite{broholm} and
show that they provide, for the first time, a direct experimental
demonstration of macroscopic entanglement in solids. The
experimental characterization of the dynamic spin correlations for
next neighbouring sites enables us to determine
concurrence~\cite{wootters} - a measure of bipartite entanglement -
and show the existence of entanglement at moderately high
temperatures (as high as 5 Kelvin). In the second, parallel,
approach we show that magnetic susceptibility at zero magnetic field
is a macroscopic thermodynamical entanglement witness for the class
of solid states systems that can be modeled by strongly alternating
spin-1/2 antiferromagnet chain. We then show that the measured
values for magnetic susceptibility of CN in 1963~\cite{berger} imply
presence of entanglement in the same temperature range (below 5
Kelvin).

CN is an accurate realization of strongly alternating
one-dimensional antiferromagnetic Heisenberg spin chain. The
corresponding spin Hamiltonian is given by
\begin{equation}
H=\sum_j (J_1 \mathbf{S}_{2j} \cdot \mathbf{S}_{2j+1} + J_2
\mathbf{S}_{2j+1} \cdot \mathbf{S}_{2j+2}),
\end{equation}
representing pairs of spins that are alternately coupled by strong
intradimer $J_1 \!\approx \! 0.44$ meV and weak interdimer $J_2
\!\approx \! 0.11$ meV coupling~\cite{bonner}. This can be described
by a model of antiferromagnetically coupled spin pairs - dimers -
which are themselves coupled by weaker antiferromagnetic
interaction. The model has a highly entangled nontrivial spin-0
ground state~\cite{barnes,tennant} and for all $0\!\leq
\!J_2/J_1\!<\!1$ has a gap of the order of $\Delta \approx J_1$ to
the first excitation, which is a band of spin-1 excitations
(magnons). However, because here $J_2/J_1 \!\approx \!
0.24$~\cite{broholm,bonner}, it is useful to think of CN as a chain
of uncoupled spin dimers. Each dimer then has a singlet ground state
and the triplets are the degenerate excited states. The existence of
the energy gap gives the first estimate for persistence of
entanglement for temperature range below $T \!\approx \!J_1/k \!
\approx \! 5 K $, where $k$ is the Boltzmann constant.

Next we describe the main experimental results of
Ref.~\cite{broholm}. We will follow the discussion given there. CN
has a monoclinic crystal structure with space group $I12/c1$ and
with low-temperature parameters $a\!=\!16.1 \AA$, $b\!=\!4.9\AA$,
$c\!=\!15.8\AA$, and $\beta=92.9^{\circ}$. The vector connecting
dimers center to center is $\mathbf{u}_0\!=\![111]/2$ for half the
chain, and $\mathbf{u'}_0\!=\![1\overline{1}1]/2$ for the other
half. The corresponding intradimer vectors are
$\mathbf{d}_1\!=\![0.252,\pm 0.027,0.228]$, respectively. In the
experiment the neutron scattering intensity was measured in the
temperature range $0.31K \!<\! T \!<\! 7.66 K$  (i.e. $0.06 J_1
\!<\!k T \!<\! 1.5 J_1)$ as a function of energy transfer $\hbar
\omega$ and wave vector transfer $\mathbf{Q}$. Count rates were
normalized to incoherent elastic scattering from the sample to
provide absolute intensity
$\tilde{I}(\mathbf{Q},\omega)=|\frac{g}{2} F(Q)|^2 2
S(\mathbf{Q},\omega)$. Here $g=\sqrt{(g^2_b+g^2_{\bot})/2}=2.22$
with $g_b\!=\!2.31$ and $g_\bot\!=\! 2.11$ that show small
anisotropy of $g$-factor along and perpendicular to the
crystallographic direction $\mathbf{b}$~\cite{bonner}, $F(Q)$ is
the magnetic form factor for Cu$^{2+}$~\cite{brown}, and
$S(\mathbf{Q},\omega)$ is the scattering function~\cite{lovesey}.

The direct link between the microscopic structure as given by the
correlation function between spins and the intensity of inelastic
neutron scattering is given by an exact sum rule (the first
$\omega$-moment of scattering cross section)~\cite{hohenberg}
\begin{eqnarray}
\hbar \langle\omega \rangle_{\mathbf Q} &\equiv& \hbar^2
\int_{-\infty}^{+\infty} S(\mathbf{Q},\omega) d\omega \nonumber \\
&=& -\frac{1}{3} \sum_\mathbf{d} J_\mathbf{d} \langle \mathbf{S}_0
\cdot \mathbf{S}_\mathbf{d} \rangle (1-\cos \mathbf{Q} \cdot
\mathbf{d}), \label{nena}
\end{eqnarray}
where $\{\mathbf d\}$ is the set of all bond vectors connecting a
spin to its neighbours, $S(\mathbf{Q},\omega)$ is single site
normalized and $\langle \mathbf{S}_0 \cdot \mathbf{S}_\mathbf{d}
\rangle \equiv \langle S^x_0 S^x_\mathbf{d} \rangle + \langle S^y_0
S^y_\mathbf{d} \rangle + \langle S^z_0 S^z_\mathbf{d} \rangle$ is
the sum over correlations for three orthogonal directions $x$, $y$
and $z$.

The intradimer correlation $\langle \mathbf{S}_0 \cdot
\mathbf{S}_{\mathbf{d}_1} \rangle$ between next neighbouring spins
was extracted from the global fits of the following
phenomenological form for $S(\mathbf{Q},\omega)$ to the complete
data set at each of temperatures (it was used more than 1000 data
points per parameter):
\begin{equation}
S(\mathbf{Q},\omega) = \frac{\hbar \langle\omega \rangle_{\mathbf
Q}}{\epsilon(\mathbf{Q})} \frac{1}{1-\exp[-\beta
\epsilon{(\mathbf{Q})}]} f[\hbar\omega - \epsilon(\mathbf{Q})].
\label{SMA}
\end{equation}
Here $\beta=1/(kT)$, $\hbar \langle\omega \rangle_{\mathbf Q}$
stands for expression (\ref{nena}), $f(E)$ is a normalized spectral
function and $\epsilon(\mathbf{Q})$ is the dispersion relation.
Equation (\ref{SMA}) represents the ''single mode approximation"
which is valid for sufficiently low
temperatures~\cite{hohenberg,girvin}. The dispersion relation is
used in the variational form based on the first order perturbation
theory~\cite{harris}
\begin{equation}
\epsilon(\mathbf{Q}) = J_1-\frac{1}{2} \sum_{\mathbf{u}}
J_{\mathbf{u}} \cos\mathbf{Q} \cdot \mathbf{u}, \label{mrs}
\end{equation}
where the vectors $\{\mathbf{u}\}$ connect neighbouring dimers
center to center both within and between the chains.

\begin{figure}
\begin{center}
\includegraphics[clip=true,angle=0,width=8.8cm]{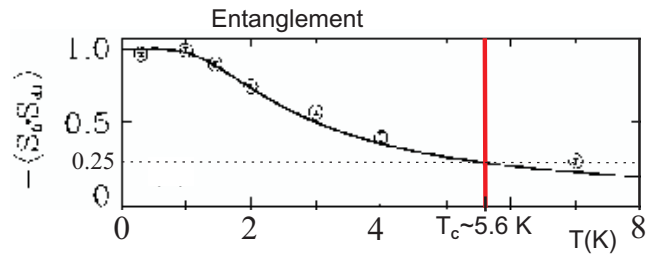}
\caption{The temperature dependence of spin correlation function
$\langle \mathbf{S}_0 \cdot \mathbf{S}_{\mathbf{d}_1} \rangle$ for
neighbouring sites in CN. The figure is taken from
Ref.~\cite{broholm}. The open circles correspond to the temperatures
at which measurements were performed, the solid black line were
obtained from fits in Ref.~\cite{broholm}. The solid red line is
from this work. It distinguishes temperature ranges with and without
entanglement in CN. The critical temperature is around
$T^{exp}_c\!\approx\!5K$ and 0.25 on the $y$-axis is the maximal
value for $\langle \mathbf{S}_0 \cdot \mathbf{S}_{\mathbf{d}_1}
\rangle$ achievable with separable states. For $T \!< \!T^{exp}_c$
concurrence is given by $C\!=\!-2\langle \mathbf{S}_0 \cdot
\mathbf{S}_{\mathbf{d}_1} \rangle -(1/2)$ and has the same
functional dependence on temperature as given in the figure. For $T
\!\geq \!T^{exp}_c$, $C$ vanishes.} \label{figure1}
\end{center}
\end{figure}

At $T=0.3K$ they used $f(E)=\delta(E)$ and obtained the global fit
with an overall factor and four exchange coupling constants
($J_1,J_2$ and the constants $J_L$ and $J_R$ for coupling between
the chains~\cite{broholm}) in Eq.~(\ref{mrs}) as the only fit
parameters. The fit gives $\langle \mathbf{S}_0 \cdot
\mathbf{S}_{\mathbf{d}_1} \rangle = -0.9(2)$ in agreement with the
value of -3/4 for dimers in the singlet state. To perform the fit at
higher temperatures the spectral function was replaced by a
normalized Gaussian with half-width-half-maximum
$\Gamma(\tilde{q})=\Gamma_0 + \frac{\Gamma_1}{2} \cos\tilde{q}$,
where $\tilde{q}=\mathbf{Q} \cdot \mathbf{u}_0$ is wave vector
transfer along the chain. Also, the dispersion relation (\ref{mrs})
was replaced by the form $\epsilon(\mathbf{Q}) = J_1-n(T)
\frac{1}{2} \sum_{\mathbf{u}} J_{\mathbf{u}} \cos\mathbf{Q} \cdot
\mathbf{u}$, where renormalization factor $n(T)$ was introduced to
account for finite temperatures. The prefactor for global fits at
each temperature yields the intradimer spin correlations $\langle
\mathbf{S}_0 \cdot \mathbf{S}_{\mathbf{d}_1} \rangle$ as given in
Fig.~\ref{figure1}. The correlations decreases with temperature
increase due to mixing of the triplets with the singlet state of
each dimer.

We will now show that the values estimated for the correlation
function can only be explained if entanglement is present in the
solid. We first show that $|\langle \mathbf{S}_0 \cdot
\mathbf{S}_{\mathbf{d}_1} \rangle |$ is an entanglement witness.
The proof is based on the fact that for any product state of a
pair of the spins one has
\begin{eqnarray}
|\langle \mathbf{S}_0 \cdot \mathbf{S}_{\mathbf{d}_1} \rangle| &=&
|\langle S^x_0 \rangle \langle S^{x}_{\mathbf{d}_1} \rangle +
\langle S^{y}_0 \rangle \langle S^{y}_{\mathbf{d}_1} \rangle +
\langle S^{z}_0 \rangle \langle S^{z}_{\mathbf{d}_1} \rangle|
\nonumber \\ &\leq & |\mathbf{S}_0| |\mathbf{S}_{\mathbf{d}_1}| \leq
1/4. \label{coldplay}
\end{eqnarray}
Here and throughout the paper we choose units to be consistent with
those of Ref.~\cite{broholm}; spin is expressed in units of 1/2;
$\hbar\!=\!1$. The upper bound was found by using the Cauchy-Schwarz
inequality and knowing that for any state $|\mathbf{S}| \equiv
\sqrt{\langle S^{x} \rangle^2 + \langle S^{y} \rangle^2 + \langle
S^{z} \rangle^2} \leq 1/2$. The proof is also valid for any convex
sum of product states of two spins (separable states): $\rho=\sum_k
w_k \rho^1_k \otimes \rho^2_k$ with $\sum_k w_k =1$. In
Fig.~\ref{figure1} the value of 1/4 for $ -\langle \mathbf{S}_0
\cdot \mathbf{S}_{\mathbf{d}_1} \rangle$ distinguishes temperatures
ranges with and without entanglement in CN. The critical temperature
is found to be $T^{exp}_c \! \approx \! 5K$. The reported error in
the correlation function ($\Delta= 0.2$ at $T=0.3K$ \cite{broholm})
implies an error of around $\Delta T \approx 1K$ in the critical
temperature.

Within the model of uncoupled dimers the isotropy of Heisenberg
interaction in spin space ensure that %$\langle
%S^\alpha_0 S^\beta_{\mathbf{d}_1} \rangle\!=\!0$ for $\alpha
%\!\neq \!\beta$ and all others are equivalent, i.e.
$\langle S^x_0 S^x_{\mathbf{d}_1} \rangle \!=\!\langle S^y_0
S^y_{\mathbf{d}_1} \rangle\!=\!\langle S^z_0 S^z_{\mathbf{d}_1}
\rangle$ and concurrence is given by $C\!=\!2\mbox{max}[0,-\langle
\mathbf{S}_0 \cdot \mathbf{S}_{\mathbf{d}_1} \rangle -(1/4)]$.
Similarly, Bell's parameter - the quantum value of Bell's
expression in the Clauser-Horne-Shymony-Holt
inequality~\cite{chsh} - is given by $(8\sqrt{2}/3) |\langle
\mathbf{S}_0 \cdot \mathbf{S}_{\mathbf{d}_1} \rangle|$. It is
higher than the local realistic limit $2$ at temperatures below
$T^{exp}_c$. Apart from rescaling both the temperature dependency
of concurrence and of Bell's parameter have the same functional
form as the correlation function in Fig.~\ref{figure1}.

The temperature dependence of $\langle \mathbf{S}_0 \cdot
\mathbf{S}_{\mathbf{d}_1} \rangle$ in Fig.~\ref{figure1} is in a
good agreement with a model of uncoupled dimers for which $\langle
\mathbf{S}_0 \cdot \mathbf{S}_{\mathbf{d}_1} \rangle \!= \! -(3/4)
\Delta n(\beta J_1)$, where $\Delta n(\beta J_1)\!=\!(1-e^{-\beta
J_1})/(1+3e^{-\beta J_1})$ is the singlet-triplet population
difference~\cite{broholm}. Within the model the theoretical
temperature dependence of concurrence is given by
$C\!=\!\mbox{max}[0, (1-3 e^{-\beta J_1})/(1+3 e^{-\beta J_1})]$,
as first obtained in Ref.~\cite{nielsen,arnesen}. The theoretical
value for the critical temperature $T^{th}_c\!=\! J_1/(k \ln
3)\!=\! 4.6 K$ is in excellent agreement with the value estimated
from the experiment.

We now proceed with our second approach. We will analyse
experimental results of a magnetic susceptibility measurement of
CN~\cite{berger} to show that the values measured at low
temperatures cannot be explained without entanglement being present
in the system. This will be based on a general proof that magnetic
susceptibility of any strongly alternating antiferromagnetic
spin-1/2 chain is a macroscopic thermodynamical entanglement
witness.

When the system is in thermal equilibrium under a certain
temperature $T$, its state is $\rho\!=\!e^{-H/kT}/Z$, where
$Z\!=\!\mbox{Tr}(e^{-H/kT})$ is the partition function and $H$ is
the Hamiltonian. The magnetic susceptibility along direction
$\alpha$ is defined as $\chi_\alpha \! \equiv \!(\partial \langle
M_\alpha \rangle /\partial B) \!=\! (g^2 \mu^2_B / kT) (\langle
(M_\alpha)^2 \rangle - \langle M_\alpha \rangle^2)$, where
$\langle M_\alpha \rangle$ is magnetization along $\alpha$,
$M_\alpha\!=\!\sum_j S^\alpha_j$, $B$ is external magnetic field,
$g$ is $g$-factor and $\mu_B$ is the Bohr magneton. Because of
the isotropy of the Hamiltonian in spin space magnetization at
zero-field vanishes for any temperature. This implies the
following form for magnetic susceptibility at zero-field:
$\chi_\alpha \!=\! (g^2 \mu^2_B / kT) \langle (M_\alpha)^2
\rangle_{B=0}\!=\!(g^2 \mu^2_B / kT) \sum_{i,j} \langle
S^\alpha_i S^\alpha_j \rangle \approx (g^2 \mu^2_B N/ kT) [(1/4)
+ \langle S^\alpha_0 S^\alpha_{\mathbf{d}_1} \rangle ]$. Here we
assume that correlations between all spins that are not nearest
neighbors are negligible compared with $\langle S^\alpha_0
S^\alpha_{\mathbf{d}_1} \rangle$. This approximation is valid for
our case of strongly alternating spin chains at low temperatures.
It is important to note that apart from the weak anisotropy in the
$g$ factor, the magnetic susceptibility at zero-field is
isotropic, i.e. $\chi_x\!=\!\chi_y\!=\!\chi_z \!\equiv \!\chi$.
Thus, if we sum the values of magnetic susceptibilities over the
three orthogonal directions $x$, $y$ and $z$ in space we obtain
$\chi\!=\! (g^2 \mu^2_B N/kT) [(1/4) + (\langle \mathbf{S}_0 \cdot
\mathbf{S}_{\mathbf{d}_1} \rangle/3)]$, where the mean value
$\langle ...\rangle$ is taken over the thermal state at
$B\!=\!0$. However, because one has $|\langle \mathbf{S}_0 \cdot
\mathbf{S}_{\mathbf{d}_1} \rangle| \leq 1/4$ for any separable
state (Eq. (\ref{coldplay})), the magnetic susceptibility for
such states is limited as given by
\begin{equation}
\chi \geq \frac{g^2 \mu^2_B N}{kT} \frac{1}{6}.
\label{macrowitness}
\end{equation}
Thus, a violation of this inequality necessarily detects
entanglement in the system.

\begin{figure}
\begin{center}
\includegraphics[clip=true, angle=0,width=9.2cm]{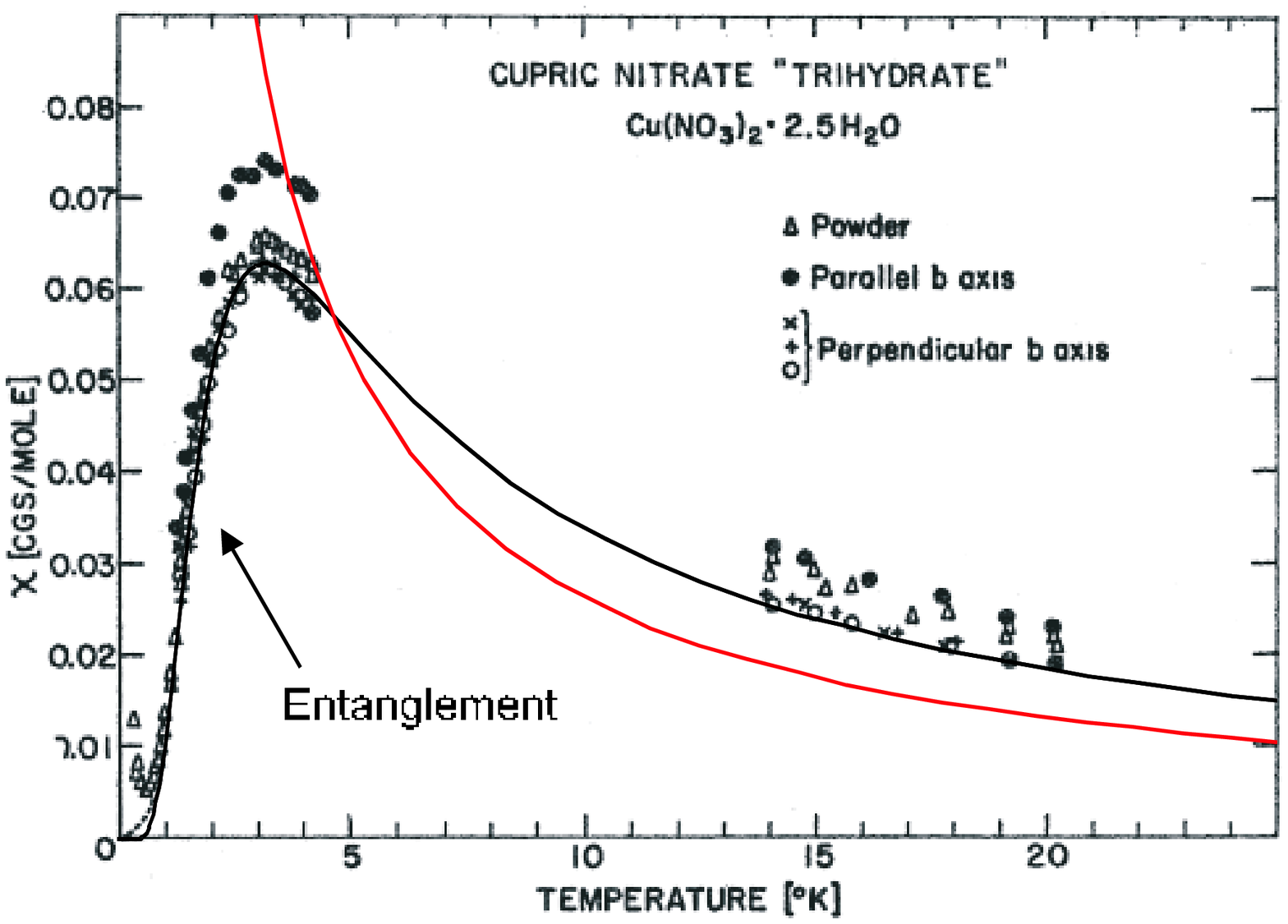}
\caption{The temperature dependence of magnetic susceptibility of
powder CN (triangles) and a single-crystal CN measured at low field
parallel (open squares) and perpendicular (open circles, crosses,
filled circles) to the monoclinic $b$ axis. The data and the figure
are from Ref.~\cite{berger}. The solid black curve is the
theoretical curve for a dimer re-scaled for the amount of noise
estimated from the experiment. This noise is computed as the ratio
of the maximal experimental value (averaged over crystal data) and
the maximal theoretical value. The solid red curve represents the
macroscopic entanglement witness~(\ref{macrowitness}) and is
re-scaled in exactly the same way. The intersection point of this
curve and the experimental one defines the temperature range (left
from the intersection point) with entanglement in CN. The critical
temperature is around $T^{exp}_c \approx 5K$. Note that the
entanglement witness will cut the experimental curve (and hence
yield a critical temperature) independently of a particular
re-scaling procedure.} \label{figure2}
\end{center}
\end{figure}

In Ref.~\cite{berger} magnetic susceptibility of CN was measured on
single crystal in $0.4-4.2 K$ and $14-20 K$ ranges of temperature.
The measurement method was based on a mutual inductance bridge
working at 275 Hz. The susceptibility was found to have a rounded
maximum at $3.2 K$ dropping very rapidly below this temperature
approaching zero at vanishing temperatures, as given in
Fig.~\ref{figure2}. Such behaviour is typical for alternating spin
chains. Thermodynamical entanglement witness~(\ref{macrowitness}) is
represented by the red solid line in Fig.~\ref{figure2}. The
measured values of magnetic susceptibility below the intersection
point of the curve representing the witness and the experimental
curve cannot be described without entanglement. The critical
temperature is around $T^{exp}_c \!\approx \!5K$. This is in
excellent agreement with the value estimated from the neutron
scattering experiment in spite the fact that different samples were
used (authors of Ref.~\cite{berger} noted possible variations of the
physical state of the sample due to high hygroscopy of CN), the two
experimental methods test entirely different physical quantities and
are almost 40 years apart (we note that no experimental error of
magnetic susceptibility measurement was reported in
Ref.~\cite{berger}).

In conclusion, we show that neutron scattering
experiments~\cite{broholm} directly and measurement of magnetic
susceptibility~\cite{berger} indirectly demonstrate presence of
macroscopic quantum entanglement in a solid (cooper nitrate). We
believe our results indicate that entanglement may play a broad
generic role in macroscopic phenomena.

The question of having macroscopic entanglement is not only
fascinating in its own right but it also has a fundamental
significance as it pushes the realm of quantum physics well into
the macroscopic world, thus opening the possibility to test
quantum theory versus alternative theories well beyond the scales
on which theirs predictions coincide. It also has important
practical implications for implementation of quantum information
processing. If the future quantum computer is supposed to reach
the stage of wide commercial application, it is likely that it
should share the same feature as the current (classical)
information technology and be based on solid state systems. It
will thus be important to derive the critical values of physical
parameters (e.g. the high-temperature limit) above which one
cannot harness quantum entanglement in solids as a resource for
quantum information processing.

We consider our work to imply that many experiments performed in the
past may still hide new and interesting physics. Experiments of
Berger {\it et al.}~\cite{berger} from 1963 and Xu {\it et
al.}~\cite{broholm} from 2000 used here are exemplary.
Interestingly, the 1963 experiment was performed long before any
serious attempts to measure entanglement begun in the seventies. It
is not even impossible that we are able to find a result from which
we can infer the existence of entanglement and which appeared well
before this concept was conceived by Schr\"{o}dinger in 1935.

%%%%%%%%%%%%%%%%%%%%%%%%%%%%%%%%%%%%%%%%%%%%%%%%%%%%%%%%%%%%%%%%%%%%%%%%%%%%%%

\end{document}